\pdfoutput=1
\documentclass[10pt,a4paper]{article}
\usepackage{cite}

\usepackage{amsmath,amsthm,amssymb}
\usepackage[pdftex]{graphicx}
\usepackage{multirow}
\usepackage{graphicx}
\usepackage[utf8]{inputenc}
\usepackage{authblk}

\title{Exploring the predictability of range-based volatility estimators using RNNs}
\author[1]{Gábor Petneházi\thanks{gabor.petnehazi@science.unideb.hu}}
\author[2]{József Gáll\thanks{gall.jozsef@inf.unideb.hu}}
\affil[1]{Doctoral School of Mathematical and Computational Sciences, University of Debrecen}
\affil[2]{Department of Applied Mathematics and Probability Theory, University of Debrecen}
\date{}

\begin{document}
\maketitle

\begin{abstract}
We investigate the predictability of several range-based stock volatility estimators, and compare them to the standard close-to-close estimator which is most commonly acknowledged as the volatility. The patterns of volatility changes are analyzed using LSTM recurrent neural networks, which are a state of the art method of sequence learning. We implement the analysis on all current constituents of the Dow Jones Industrial Average index, and report averaged evaluation results. We find that changes in the values of range-based estimators are more predictable than that of the estimator using daily closing values only.
\end{abstract}

\section{Motivation}
The volatility of assets has an important role in several areas of finance. As a measure of riskiness, it is a key factor in, for example, portfolio management and option pricing. A good understanding of the nature and evolution of return volatilities is obviously valuable for financial practitioners.\\
Volatility quantifies the dispersion of returns. Unfortunately, this dispersion can not be measured --- volatility is not directly observable. Hence we need to estimate it, with not having a reliable benchmark.\\
Several studies have tried to explore and understand the nature of this unknown volatility. One reasonable approach is sampling from the price process frequently, so that we do not lose too much data. \cite{andersen2001distribution} analyzed the properties of stock market volatility using five-minute returns. They report that daily variances significantly fluctuate through time, and their distributions are extremely right-skewed and leptokurtic, while logarithmic standard deviations approximate the normal distribution well. \cite{engle2007good} outlined several stylised facts of volatility that have emerged in previous studies. Persistence: large moves are usually followed by large moves, and small moves are usually followed by small moves in the price process. Mean reversion: usually there’s a normal level of volatility to which it returns after uplifts and falls. Asymmetric impact of innovations: positive and negative shocks have different impacts on volatility. Influence of exogenous variables: information outside the price series (e.g., announcements) could have an impact on volatility.\\
Those features suggest that if we could measure volatility, it should be somewhat forecastable. But we can not measure it --- the best we can do is coming up with reasonable proxies.\\
One such proxy is the standard deviation of returns --- returns, which are usually calculated from daily closing prices. It is obvious that by sampling the asset’s price more frequently, we could make better estimates of its unobservable true volatility. If, for example, we measured the daily price ranges (i.e.\ daily high minus daily low), we would already know a lot more about the unseen path of the prices.\\
Unlike high-frequency (say, minutely) data, daily open, high, low and close values are freely available. Finding good estimators that use these 4 daily values only is therefore a challenging and important task.\\
It this paper we are going to compare various range-based volatility estimators according to their predictability. We argue that those estimators whose changes are easier to predict, can be more useful in practice. Forecasts can move historical volatilities a bit forward into the future, and knowing something about the future is valuable.\\

\section{Volatility estimators}
Volatility is most often calculated simply as the standard deviation of returns (\ref{eq:close-to-close}). In the formula below, $C_t$ is the closing price of day $t$, and $N$ is the number of days used in the calculation. As volatility should measure the dispersion of the prices, standard deviation is a very reasonable proxy.

\begin{equation} \label{eq:close-to-close}
\sigma = \sqrt{F}\sqrt{\frac{\sum_{t=1}^{N}(ln(\frac{C_t}{C_{t-1}})-\overline{ln(\frac{C_t}{C_{t-1}})})^{2}}{N-1}}
\end{equation}

However, when returns are calculated on a daily basis (as the difference of log closing prices), this simple and intuitive formula ignores all intraday price movements, which is a great loss of information.\\
The so-called range-based volatility estimators use daily open, high, low and close values to make volatility estimates. Several such formulas has been proposed in the history of volatility estimation. Here we are going to present some of the better known range-based volatility formulas. $O_t$, $H_t$, $L_t$, and $C_t$ stand for the open, high, low, and close price at time $t$, respectively. $N$, again, is the size of the time window in days for calculating the volatilities, while $F$ is just for scaling the results to another time unit.\\
\cite{parkinson1980extreme} proposed the extreme value method for variance estimation (\ref{eq:parkinson}). He showed that using high and low prices, we may get an estimate that is far superior to the standard close-to-close formula.

\begin{equation} \label{eq:parkinson}
\sigma_{P} = \sqrt{F}\sqrt{\frac{\frac{1}{4\, ln(2)}\sum_{t=1}^{N}(ln(\frac{H_t}{L_t}))^{2}}{N}}
\end{equation}

\cite{garman1980estimation} published estimators using open, high, low and close values (\ref{eq:garmanklass}). Their results demonstrate much higher efficiency factors than that of the close-to-close estimator.

\begin{equation} \label{eq:garmanklass}
\sigma_{GK} = \sqrt{F}\sqrt{\frac{\sum_{t=1}^{N}0.5(ln(\frac{H_t}{L_t}))^2-(2\, ln(2)-1)(ln(\frac{C_t}{O_t}))^2}{N}}
\end{equation}

The Parkinson and Garman-Klass volatility estimators assume the asset prices follow a continuous Brownian motion with no drift. \cite{rogers1991estimating} proposed a formula that allows for drifts (\ref{eq:rogerssatchell}). \cite{rogers1994estimating} investigated the efficiency of volatility estimators through simulation, and found that the Rogers-Satchell method is superior to the Garman-Klass if there is a time-varying drift in the data. However, when there’s no drift, Garman-Klass outperforms Rogers-Satchell, so the former should be preferred when the expected returns are less volatile.

\begin{equation} \label{eq:rogerssatchell}
\sigma_{RS} = \sqrt{F}\sqrt{\frac{\sum_{t=1}^{N}ln(\frac{H_t}{O_t})(ln(\frac{H_t}{O_t})-ln(\frac{C_t}{O_t}))+ln(\frac{L_t}{O_t})(ln(\frac{L_t}{O_t})-ln(\frac{C_t}{O_t}))}{N}}
\end{equation}

\cite{yang2000drift} published a formula which is unbiased, drift-independent, and consistent in dealing with opening jumps (\ref{eq:yangzhang}). This latter feature is unique among the examined formulas.

\begin{equation} \label{eq:yangzhang}
\begin{gathered}
\sigma_{YZ} = \\ \sqrt{F} \sqrt{\frac{\sum_{t=1}^{N}(ln(\frac{O_t}{C_{t-1}})-\overline{ln(\frac{O_t}{C_{t-1}})})^2}{N-1}+\frac{k\sum_{t=1}^{N}(ln(\frac{C_t}{O_{t}})-\overline{ln(\frac{C_t}{O_{t}})})^2}{N-1}+(1-k)V_{RS}} \\
k=\frac{0.34}{1.34+\frac{N+1}{N-1}} \\
V_{RS} = \frac{\sum_{t=1}^{N}ln(\frac{H_t}{O_t})(ln(\frac{H_t}{O_t})-ln(\frac{C_t}{O_t}))+ln(\frac{L_t}{O_t})(ln(\frac{L_t}{O_t})-ln(\frac{C_t}{O_t}))}{N}
\end{gathered}
\end{equation}

Range-based volatility estimation has quite a long history and evolution. Here we have only mentioned the formulas that we are going to use in this work. \cite{chou2010range} gives a detailed review of the development of range-based volatility estimators.
\begin{figure}
\centering
\includegraphics[width=\textwidth,height=\textheight,keepaspectratio]{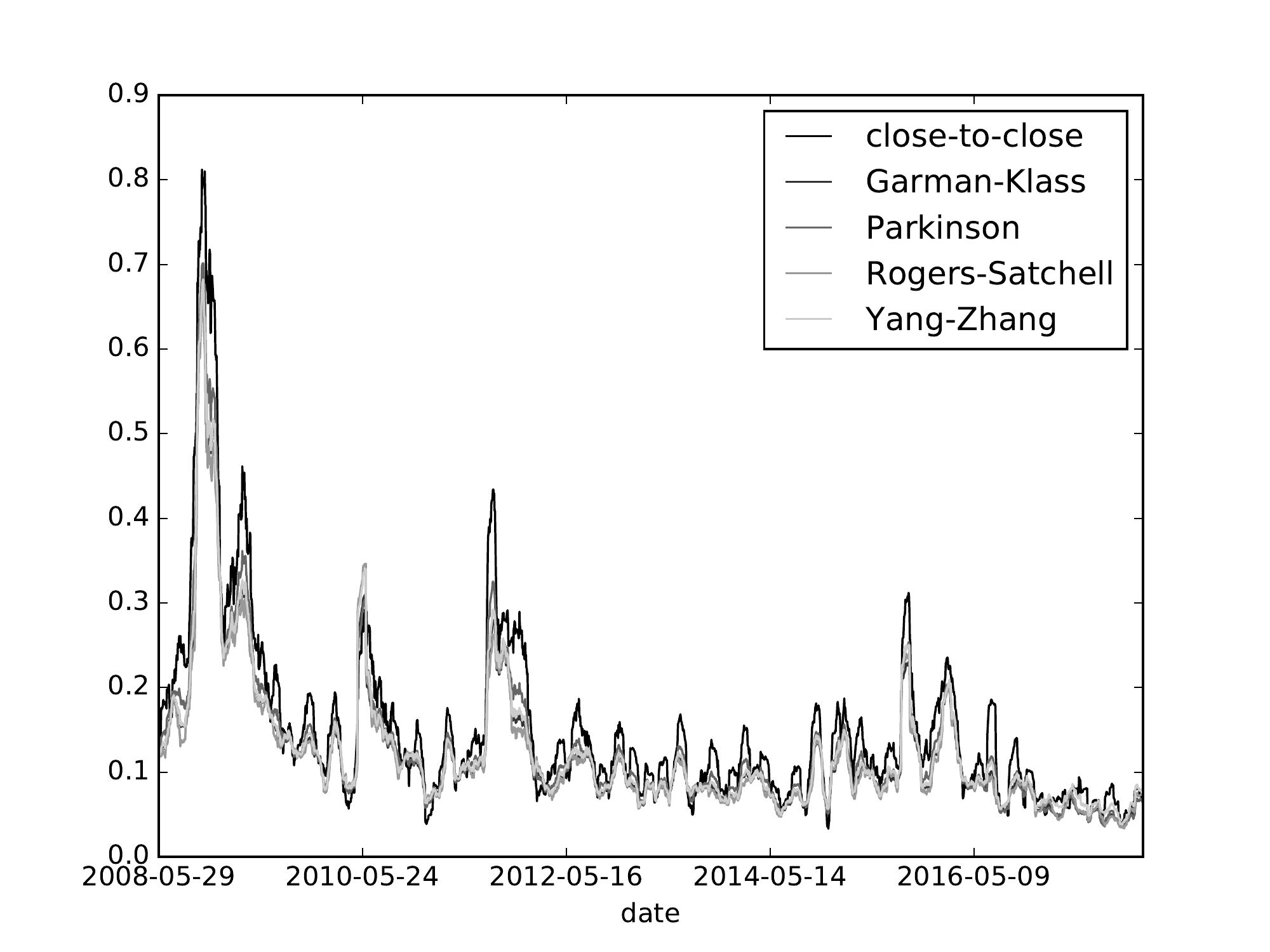}
\caption[Volatility estimates for the DJIA]{Volatility estimates for the DJIA index in the observed period}
\end{figure}
\begin{table}[h!]
\centering
\begin{tabular}{ |c|c|c|c|c|c|c|c|c| }
 \hline
 correlations & $\sigma$ & $\sigma_{GK}$ & $\sigma_{P}$ & $\sigma_{RS}$ & $\sigma_{YZ}$ \\
 \hline
 $\sigma$&1.000&0.973&0.988&0.956&0.973 \\
 $\sigma_{GK}$&0.973&1.000&0.996&0.997&0.999 \\
 $\sigma_{P}$&0.988&0.996&1.000&0.988&0.995 \\
 $\sigma_{RS}$&0.956&0.997&0.988&1.000&0.997 \\
 $\sigma_{YZ}$&0.973&0.999&0.995&0.997&1.000 \\
 \hline
\end{tabular}
\caption{Correlations of different esimates of DJIA volatility}
\label{table:correlations}
\end{table}

All of those range-based estimators assume that the asset price follows a continuous geometric Brownian motion. This is a strict assumption. \cite{shu2006testing} analyzed all 4 range-based estimators that we investigate, in an attempt to measure the degree to which they can be useful in real markets that deviate from the geometric Brownian motion. They found that estimates from range-based models are quite close to integrated variances computed from intraday returns with much higher computational requirements. Using simulations, Shu and Zhang also confirmed the expectation that when there is just a small drift and no opening jumps in the prices, all 4 estimators provide good estimates. When the drift is large, the Parkinson and Garman-Klass estimators overestimate the true variances, while the other two behave properly. Large opening jumps can only be handled by the Yang-Zhang estimator, all the other formulas give downward biased estimates in case of opening jumps.\\
We’ve just got the gist of the behavior of some available volatility estimators. Yet, since we can never know the true variances, it’s pretty hard to assess each formula’s exact accuracy and usefulness in real practice. We argue that a measure of forecastability could be useful for evaluating and comparing these volatility proxies.\\

\section{Volatility forecasting}
While forecasting changes in stock returns is a very hard task, forecasting the size of changes (i.e.\ the volatility) seems more promising. It also has a high importance in financial practice, and it has already been the subject of many researches.\\
\cite{poon2003forecasting} give a detailed review on 93 papers that study the forecasting power of volatility models. \cite{poon2005practical} provide a summary of the findings. They survey historical volatility, ARCH, stochastic volatility and option-implied volatility models.\\
For modeling the changes in volatility, artificial neural networks also seem a natural choice. Neural networks, even with a single hidden layer, are universal approximators \cite{hornik1991approximation}, and it makes them a strong competitor of traditional learning algorithms and time series methods.\\
\cite{malliaris1996using} forecasted implied volatilities using neural networks trained on past volatilities and other options market factors. \cite{donaldson1996forecast} used ANNs to combine different time series forecasts of stock market volatilities, and concluded that combining forecasts using ANNs generally outperforms traditional combining methods due to its flexibility. \cite{roh2007forecasting} proposed new hybrid models combining neural networks and time series models for improving volatility predictions in terms of deviation and direction accuracy.\\
Recurrent neural networks have also been applied to volatility forecasting. \cite{xiong2015deep} used LSTM networks on Google Domestic Trends data to forecast S\&P 500 volatilities. Their article is similar to ours, since we too apply LSTM RNNs to forecast range-based volatility. However, they used external data while we only use historical stock prices, and they made predictions for the daily values of the volatility estimates, while we aim for predicting the directions of daily changes.\\
Making reasonable forecasts about volatility changes can potentially help one to make profitable trading decisions. \cite{tino2000temporal}, for example, used volatility forecasts to buy or sell straddles, except when their model was uncertain about the sign of volatility change. A similar strategy is applied by \cite{dunis2002forecasting}, using RNNs for forecasting volatilities. In this article, we are going to explore the forecastability of the directions of volatility changes, which could provide some information on which positions to take.\\

\section{Recurrent neural networks}
Recurrent neural networks are neural networks for sequential data --- instead of relying on a single data point, recurrent neural networks take into account a whole sequence. An RNN (\ref{eq:vanillarnn}) uses the most recent observation together with the past to make a good decision. This trait makes them a reasonable choice for modeling the behavior of time series.

\begin{equation} \label{eq:vanillarnn}
h_{t} = \tanh \left ( W_{h}x_{t} + U_{h}h_{t-1} + b_{h} \right ) 
\end{equation}

Yet, plain RNN models (e.g., \cite{elman1990finding}) suffer from the vanishing gradient problem and are hard to train \cite{hochreiter2001gradient}. They are unable to model long-term dependencies in the data. Luckily, there are some more advanced RNN architectures that solve this problem on the expense of some model complexity. One such architecture is long short-term memory (LSTM).\\
LSTMs were invented by \cite{hochreiter1997long}. An LSTM cell gives memory to the RNN, and the ability to read, write and forget data. The cell uses separate gating units to operate these memory management abilities.

\begin{equation} \label{eq:lstm_inputgate}
i_{t} = sigmoid \left ( W_{i}x_{t} + U_{i}h_{t-1} + b_{i} \right )
\end{equation}
\begin{equation} \label{eq:lstm_forgetgate}
f_{t} = sigmoid \left ( W_{f}x_{t} + U_{f}h_{t-1} + b_{f} \right ) 
\end{equation}
\begin{equation} \label{eq:lstm_outputgate}
o_{t} = sigmoid \left ( W_{o}x_{t} + U_{o}h_{t-1} + b_{o} \right )
\end{equation}
\begin{equation} \label{eq:lstm_cellstate}
c_{t} = f_{t} \odot c_{t-1} + i_{t} \odot tanh \left ( W_{c}x_{t} + U_{c}h_{t-1} + b_{c} \right )
\end{equation}
\begin{equation} \label{eq:lstm_hiddenstate}
h_{t} = o_{t} \odot tanh \left ( c_{t} \right )
\end{equation}

This looks a bit complicated, but it is just as easy to train as any other neural network. It learns what it needs to learn, and forgets what it needs to forget --- this makes LSTM well suited for analyzing long time series data.\\
$x_{t} \in \mathbb{R}^n$ is the input to the LSTM cell. $h_{t} \in \mathbb{R}^h$ denotes the output from the LSTM, which is usually called the hidden state (\ref{eq:lstm_hiddenstate}). $c_{t} \in \mathbb{R}^h$ is the so-called cell state (\ref{eq:lstm_cellstate}), which represents the memory. $i_{t} \in \mathbb{R}^h$, $f_{t} \in \mathbb{R}^h$, and $o_{t} \in \mathbb{R}^h$ are the input, forget and output gates. The input gate (\ref{eq:lstm_inputgate}) calculates what to keep in memory, the forget gate (\ref{eq:lstm_forgetgate}) calculates what to remember, the output gate (\ref{eq:lstm_outputgate}) calculates which part of the memory to use immediately. They do these things by applying some simple mathematical operations on the input data, the previous hidden state, and the corresponding learnable weights $W \in \mathbb{R}^{h \times n}$ and $U \in \mathbb{R}^{h \times h}$, which can easily be optimized using backpropagation through time (e.g., \cite{greff2017lstm}).\\
The LSTM formulas are rather formidable at first sight, but they form a system that is fairly intuitive, and works, in many cases, amazingly well.\\
Some recent applications of RNNs to time series forecasting are, e.g., \cite{che2016recurrent}, \cite{cinar2017time}, \cite{cinar2017position}, \cite{hsu2017time}, \cite{laptev2017time}. LSTMs are also used in some financial studies for modeling time series data like historical volatility. This study aims to contribute to this research area by comparing the predictability of range-based volatility estimates using LSTM networks.\\

\section{Data}
Our dataset was obtained from Yahoo Finance (https://finance.yahoo.com/). We have downloaded 10 years (from 2008-01-01 to 2017-12-31) of daily open, high, low and close values for all current constituents of the Dow Jones Industrial Average index. (1 out of the 30 stocks is missing the first few months' data since its IPO took place in March 2008.)\\
We used the previously presented formulas for quantifying volatility, namely, the close-to-close, the Garman-Klass, the Parkinson, the Rogers-Satchell, and the Yang-Zhang estimator. All volatility estimates were calculated using a window of 21 days.\\
We used very few data for training our neural networks. The exact values of the estimates has been dropped, and we only kept a single binary variable indicating the direction of daily changes --- 1s for upward movements, 0s everywhere else.\\
We used the first 70\% of the available data for training the LSTM models, and we made one-day-ahead forecasts on the remaining 30\%, which is roughly the last 3 years.\\

\section{Neural Network Architecture}
Our LSTM recurrent neural network was built in Keras \cite{chollet2015keras} with TensorFlow cite{tensorflow2015-whitepaper} backend.\\
We used a 2-layer RNN with 10 hidden units in each layer. We chose the Adam optimizer \cite{kingma2014adam} to minimize the loss function, in this case, binary cross entropy. A dropout \cite{srivastava2014dropout} of .3 was applied on the non-recurrent connections. The learning rate was set to .001.\\
The series of volatility directions were unrolled for 10 days, and the unrolled subsequences were fed to the algorithm in batches of 32.\\
All experiments were run for no more than 300 epochs with the following early stopping rule: the training stops when it fails to improve the validation loss for 50 epochs in a row.\\
No thorough hyperparameter optimization was conducted. We only aimed to find a reasonable setting which is appropriate for comparing the predictability of our examined volatility formulas.\\

\section{Results}
We used roughly the last 3 years of our 10-year dataset for out-of-sample validation. 4 evaluation metrics are reported: accuracy, precision, recall and F1 score. Those metrics were averaged over all 30 constituents of the Dow Jones Industrial Average stock market index. For each stock, we have trained an individual RNN model.

\begin{table}[h!]
\centering
\begin{tabular}{ |c|c|c|c|c|c|c|c|c| }
 \hline
 \multirow{2}{*}{} & \multicolumn{2}{|c|}{Accuracy} & \multicolumn{2}{|c|}{Precision} & \multicolumn{2}{|c|}{Recall} & \multicolumn{2}{|c|}{F1} \\
 \cline{2-9}
 & mean & std & mean & std & mean & std & mean & std \\
 \hline
 close-to-close&0.51&0.02&0.52&0.07&0.31&0.25&0.33&0.17 \\
 Garman-Klass&\textbf{0.57}&0.03&\textbf{0.63}&0.05&0.30&0.10&0.40&0.10 \\
 Parkinson&\textbf{0.57}&0.02&\textbf{0.63}&0.04&\textbf{0.32}&0.09&\textbf{0.42}&0.08 \\
 Rogers-Satchell&0.55&0.03&0.61&0.04&0.26&0.10&0.35&0.10 \\
 Yang-Zhang&\textbf{0.57}&0.02&\textbf{0.63}&0.05&0.29&0.09&0.39&0.08 \\
 \hline
\end{tabular}
\caption{Evaluation metrics for one-day-ahead direction-of-change forecasts}
\label{table:results_1}
\end{table}

Table \ref{table:results_1} displays our results for this experiment. While the close-to-close estimator’s accuracy was barely higher than 50\%, neither of the range-based volatilities’ averaged accuracies was below 55\%. It seems to be a considerable difference.\\
Despite the promising accuracies, the F1 score, being the harmonic mean of precision and recall, was consistently below .5 in each case.\\
Precision is the fraction of our predicted upward movements that really were increases in the volatility. Recall is the fraction of real upward movements that we’ve predicted to be so. While each estimator’s precision was higher than the overall accuracy, the recall was very poor. It means that our algorithm struggled in finding the upward movements. It simply follows from the fact that the RNN has chosen to predict downward movements in a higher proportion. Since the ratio of upward movements was usually close to .5, this behavior of the algorithm does not invalidate its prediction ability. Yet, it is obviously not a preferred property, especially when we assume that identifying a rise in the volatility is more valuable than identifying a drop.\\
To solve the issue of low recall, we tried lowering the threshold of making an upward guess from the default of .5.

\begin{table}[h!]
\centering
\begin{tabular}{ |c|c|c|c|c|c|c|c|c| }
 \hline
 \multirow{2}{*}{} & \multicolumn{2}{|c|}{Accuracy} & \multicolumn{2}{|c|}{Precision} & \multicolumn{2}{|c|}{Recall} & \multicolumn{2}{|c|}{F1} \\
 \cline{2-9}
 & mean & std & mean & std & mean & std & mean & std \\
 \hline
 close-to-close&0.50&0.02&0.50&0.01&\textbf{0.88}&0.16&\textbf{0.63}&0.06 \\
 Garman-Klass&\textbf{0.58}&0.03&0.59&0.03&0.49&0.09&0.53&0.06 \\
 Parkinson&\textbf{0.58}&0.02&\textbf{0.60}&0.03&0.46&0.09&0.51&0.06 \\
 Rogers-Satchell&0.57&0.02&0.58&0.03&0.45&0.10&0.50&0.07 \\
 Yang-Zhang&\textbf{0.58}&0.02&0.59&0.03&0.47&0.10&0.52&0.07 \\
 \hline
\end{tabular}
\caption{Evaluation metrics for predictions with .45 probability threshold}
\label{table:results_2}
\end{table}

Table \ref{table:results_2} displays our results using a threshold of .45. In this case, we force the RNN to predict more increases. The recall values increased, leading to above .5 F1 scores, while not sacrificing the overall accuracy. In fact, the accuracy increased a bit for all range-based estimators.\\
Finally, we freed the algorithm from having to make predictions at all times. It may be preferable to let the algorithm decide if it has the necessary confidence to make a prediction. We chose to flag prediction probabilities between .4 and .5 as unconfident, and only kept the days with estimated probabilities outside this range. (All those thresholds were chosen arbitrarily.)

\begin{table}[h!]
\centering
\begin{tabular}{ |c|c|c|c|c|c|c|c|c| }
 \hline
 \multirow{2}{*}{} & \multicolumn{2}{|c|}{Accuracy} & \multicolumn{2}{|c|}{Precision} & \multicolumn{2}{|c|}{Recall} & \multicolumn{2}{|c|}{F1}\\
 \cline{2-9}
 & mean & std & mean & std & mean & std & mean & std \\
 \hline
 close-to-close&0.51&0.11&0.51&0.11&\textbf{0.93}&0.20&\textbf{0.66}&0.13 \\
 Garman-Klass&0.61&0.03&\textbf{0.63}&0.04&0.50&0.11&0.55&0.07 \\
 Parkinson&0.61&0.03&\textbf{0.63}&0.05&0.45&0.11&0.51&0.08 \\
 Rogers-Satchell&0.59&0.03&0.62&0.05&0.46&0.15&0.52&0.12 \\
 Yang-Zhang&\textbf{0.62}&0.03&\textbf{0.63}&0.04&0.48&0.12&0.54&0.09 \\
 \hline
\end{tabular}
\caption{Evaluation metrics for confident predictions (P\textgreater .5 or P\textless .4)}
\label{table:results_3}
\end{table}

Table \ref{table:results_3} presents the evaluation metrics for the confident predictions. By dropping (quite) some uncertain predictions, we have exceeded 60\% accuracy with 3 out of 4 range-based estimators.

\begin{table}[h!]
\centering
\begin{tabular}{ |c|c|c|c|c| }
 \hline
 close-to-close&Garman-Klass&Parkinson&Rogers-Satchell&Yang-Zhang \\
 \hline 
 0.28&0.66&0.67&0.54&0.64
 \\
 \hline
\end{tabular}
\caption{Average proportions of confidently predicted directions}
\label{table:keep_ratios}
\end{table}

Table \ref{table:keep_ratios} displays the proportions of predictions that remained after excluding the uncertain ones. It seems that estimators with lower accuracies has more predictions close to the .45 binary decision threshold. Hence weak forecasts make less guesses, which is preferable.\\
In this experiment, all four range-based estimators clearly outperformed the benchmark close-to-close estimator in terms of predictability. The range-based estimators generated similar results, though the Rogers-Satchell estimator performed slightly worse than the other three.\\
All of those volatility estimators move closely together, as expected, having above .95 correlations for the Dow Jones Industrial Average index (Table \ref{table:correlations}). We could observe similary high correlation in case of the individual components as well. It is therefore quite remarkable, that while the close-to-close estimator seems essentially unpredictable, the directional changes of range-based estimators were so easily detected from so few data to near 60\% accuracy, that it calls for further research.\\

\section{Conclusions}
We can conclude that the movements of range-based volatility calculations can be forecasted to some degree, using long short-term memory recurrent neural networks and very little data --- only historical patterns of up and down movements.\\
There’s not much difference in the predictability of the range-based volatility estimators, however they all seem to be easier to forecast than the baseline close-to-close estimator, which is most commonly used and acknowledged as financial volatility.\\

\bibliography{article_refs.bib}{}

\begin{thebibliography}{}

\bibitem[Andersen et~al., 2001]{andersen2001distribution}
Andersen, T.~G., Bollerslev, T., Diebold, F.~X., and Ebens, H. (2001).
\newblock The distribution of realized stock return volatility.
\newblock {\em Journal of financial economics}, 61(1):43--76.

\bibitem[Che et~al., 2016]{che2016recurrent}
Che, Z., Purushotham, S., Cho, K., Sontag, D., and Liu, Y. (2016).
\newblock Recurrent neural networks for multivariate time series with missing
  values.
\newblock {\em arXiv preprint arXiv:1606.01865}.

\bibitem[Chollet et~al., 2015]{chollet2015keras}
Chollet, F. et~al. (2015).
\newblock Keras.

\bibitem[Chou et~al., 2010]{chou2010range}
Chou, R.~Y., Chou, H., and Liu, N. (2010).
\newblock Range volatility models and their applications in finance.
\newblock In {\em Handbook of quantitative finance and risk management}, pages
  1273--1281. Springer.

\bibitem[Cinar et~al., 2017a]{cinar2017position}
Cinar, Y.~G., Mirisaee, H., Goswami, P., Gaussier, E., A{\"\i}t-Bachir, A., and
  Strijov, V. (2017a).
\newblock Position-based content attention for time series forecasting with
  sequence-to-sequence rnns.
\newblock In {\em International Conference on Neural Information Processing},
  pages 533--544. Springer.

\bibitem[Cinar et~al., 2017b]{cinar2017time}
Cinar, Y.~G., Mirisaee, H., Goswami, P., Gaussier, E., Ait-Bachir, A., and
  Strijov, V. (2017b).
\newblock Time series forecasting using rnns: an extended attention mechanism
  to model periods and handle missing values.
\newblock {\em arXiv preprint arXiv:1703.10089}.

\bibitem[Donaldson and Kamstra, 1996]{donaldson1996forecast}
Donaldson, R.~G. and Kamstra, M. (1996).
\newblock Forecast combining with neural networks.
\newblock {\em Journal of Forecasting}, 15(1):49--61.

\bibitem[Dunis and Huang, 2002]{dunis2002forecasting}
Dunis, C.~L. and Huang, X. (2002).
\newblock Forecasting and trading currency volatility: An application of
  recurrent neural regression and model combination.
\newblock {\em Journal of Forecasting}, 21(5):317--354.

\bibitem[Elman, 1990]{elman1990finding}
Elman, J.~L. (1990).
\newblock Finding structure in time.
\newblock {\em Cognitive science}, 14(2):179--211.

\bibitem[Engle and Patton, 2007]{engle2007good}
Engle, R.~F. and Patton, A.~J. (2007).
\newblock What good is a volatility model?
\newblock In {\em Forecasting Volatility in the Financial Markets (Third
  Edition)}, pages 47--63. Elsevier.

\bibitem[Garman and Klass, 1980]{garman1980estimation}
Garman, M.~B. and Klass, M.~J. (1980).
\newblock On the estimation of security price volatilities from historical
  data.
\newblock {\em Journal of business}, pages 67--78.

\bibitem[Greff et~al., 2017]{greff2017lstm}
Greff, K., Srivastava, R.~K., Koutn{\'\i}k, J., Steunebrink, B.~R., and
  Schmidhuber, J. (2017).
\newblock Lstm: A search space odyssey.
\newblock {\em IEEE transactions on neural networks and learning systems},
  28(10):2222--2232.

\bibitem[Hochreiter et~al., 2001]{hochreiter2001gradient}
Hochreiter, S., Bengio, Y., Frasconi, P., Schmidhuber, J., et~al. (2001).
\newblock Gradient flow in recurrent nets: the difficulty of learning long-term
  dependencies.

\bibitem[Hochreiter and Schmidhuber, 1997]{hochreiter1997long}
Hochreiter, S. and Schmidhuber, J. (1997).
\newblock Long short-term memory.
\newblock {\em Neural computation}, 9(8):1735--1780.

\bibitem[Hornik, 1991]{hornik1991approximation}
Hornik, K. (1991).
\newblock Approximation capabilities of multilayer feedforward networks.
\newblock {\em Neural networks}, 4(2):251--257.

\bibitem[Hsu, 2017]{hsu2017time}
Hsu, D. (2017).
\newblock Time series forecasting based on augmented long short-term memory.
\newblock {\em arXiv preprint arXiv:1707.00666}.

\bibitem[Kingma and Ba, 2014]{kingma2014adam}
Kingma, D.~P. and Ba, J. (2014).
\newblock Adam: A method for stochastic optimization.
\newblock {\em arXiv preprint arXiv:1412.6980}.

\bibitem[Laptev et~al., 2017]{laptev2017time}
Laptev, N., Yosinski, J., Li, L.~E., and Smyl, S. (2017).
\newblock Time-series extreme event forecasting with neural networks at uber.
\newblock In {\em International Conference on Machine Learning}.

\bibitem[Malliaris and Salchenberger, 1996]{malliaris1996using}
Malliaris, M. and Salchenberger, L. (1996).
\newblock Using neural networks to forecast the s\&p 100 implied volatility.
\newblock {\em Neurocomputing}, 10(2):183--195.

\bibitem[Parkinson, 1980]{parkinson1980extreme}
Parkinson, M. (1980).
\newblock The extreme value method for estimating the variance of the rate of
  return.
\newblock {\em Journal of business}, pages 61--65.

\bibitem[Poon and Granger, 2005]{poon2005practical}
Poon, S.-H. and Granger, C. (2005).
\newblock Practical issues in forecasting volatility.
\newblock {\em Financial analysts journal}, 61(1):45--56.

\bibitem[Poon and Granger, 2003]{poon2003forecasting}
Poon, S.-H. and Granger, C.~W. (2003).
\newblock Forecasting volatility in financial markets: A review.
\newblock {\em Journal of economic literature}, 41(2):478--539.

\bibitem[Rogers et~al., 1994]{rogers1994estimating}
Rogers, L.~C., Satchell, S.~E., and Yoon, Y. (1994).
\newblock Estimating the volatility of stock prices: a comparison of methods
  that use high and low prices.
\newblock {\em Applied Financial Economics}, 4(3):241--247.

\bibitem[Rogers and Satchell, 1991]{rogers1991estimating}
Rogers, L. C.~G. and Satchell, S.~E. (1991).
\newblock Estimating variance from high, low and closing prices.
\newblock {\em The Annals of Applied Probability}, pages 504--512.

\bibitem[Roh, 2007]{roh2007forecasting}
Roh, T.~H. (2007).
\newblock Forecasting the volatility of stock price index.
\newblock {\em Expert Systems with Applications}, 33(4):916--922.

\bibitem[Shu and Zhang, 2006]{shu2006testing}
Shu, J. and Zhang, J.~E. (2006).
\newblock Testing range estimators of historical volatility.
\newblock {\em Journal of Futures Markets}, 26(3):297--313.

\bibitem[Srivastava et~al., 2014]{srivastava2014dropout}
Srivastava, N., Hinton, G., Krizhevsky, A., Sutskever, I., and Salakhutdinov,
  R. (2014).
\newblock Dropout: A simple way to prevent neural networks from overfitting.
\newblock {\em The Journal of Machine Learning Research}, 15(1):1929--1958.

\bibitem[Tino et~al., 2000]{tino2000temporal}
Tino, P., Schittenkopf, C., and Dorffner, G. (2000).
\newblock Temporal pattern recognition in noisy non-stationary time series
  based on quantization into symbolic streams. lessons learned from financial
  volatility trading.

\bibitem[Xiong et~al., 2015]{xiong2015deep}
Xiong, R., Nichols, E.~P., and Shen, Y. (2015).
\newblock Deep learning stock volatility with google domestic trends.
\newblock {\em arXiv preprint arXiv:1512.04916}.

\bibitem[Yang and Zhang, 2000]{yang2000drift}
Yang, D. and Zhang, Q. (2000).
\newblock Drift-independent volatility estimation based on high, low, open, and
  close prices.
\newblock {\em The Journal of Business}, 73(3):477--492.

\end{thebibliography}
\bibliographystyle{apalike}

\end{document}